# Bayesian Inference for Gaussian Mixed Graph Models

**Ricardo Silva**
Gatsby Computational Neuroscience Unit
University College London
rbas@gatsby.ucl.ac.uk

**Zoubin Ghahramani**
Department of Engineering
University of Cambridge
zoubin@eng.cam.ac.uk

## Abstract

We introduce priors and algorithms to perform Bayesian inference in Gaussian models defined by acyclic directed mixed graphs. Such a class of graphs, composed of directed and bi-directed edges, is a representation of conditional independencies that is closed under marginalization and arises naturally from causal models which allow for unmeasured confounding. Monte Carlo methods and a variational approximation for such models are presented. Our algorithms for Bayesian inference allow the evaluation of posterior distributions for several quantities of interest, including causal effects that are not identifiable from data alone but could otherwise be inferred where informative prior knowledge about confounding is available.

## 1 CONTRIBUTION

Directed mixed graphs (DMGs) are graphs with directed and bi-directed edges. In particular, acyclic directed mixed graphs have no directed cycle, i.e., no sequence of directed edges $X \rightarrow \cdots \rightarrow X$ that starts and ends on the same node. Such a representation encodes a set of conditional independencies among random variables, which can be read out of a graph by using a criterion known as m-separation, a natural extension of the d-separation criterion used for directed acyclic graphs (DAGs) (Richardson, 2003).

In a DMG, two adjacent nodes might be connected by up to two edges, where in this case one has to be bi-directed and the other directed. Figure 1 illustrates a simple case. The appeal of this graphical family lies on the representation of the marginal independence structure among a set of observed variables, assuming they are part of a larger DAG structure that includes hidden variables. This is an important feature in, e.g., causal models that are robust to unmeasured confounding. In the example of Figure 1, the direct causal effect of $Y_2$ on $Y_3$ can be in principle separated from the association due to unspecified hidden common causes represented by $Y_2 \leftrightarrow Y_3$.

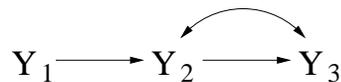

Figure 1: A directed mixed graph.

DMGs form a family of independence models that is closed under marginalization. In general, however, a family of distributions that respects the encoded independence constraints will not be closed.

This paper will focus on Gaussian models defined on parameterized *acyclic* DMGs (ADMGs). Our contribution is, to the best of our knowledge, the first fully Bayesian treatment of this family of models. This provides a Bayesian alternative to, e.g., the maximum likelihood estimator of Drton and Richardson (2004). We will define priors and describe novel algorithms for Bayesian inference in Gaussian acyclic DMG models.

The paper is organized as follows: Section 2 is a brief review of Gaussian DMGs. Section 3 defines our priors. Section 4 describes a Monte Carlo procedure for the special case of Gaussian bi-directed models. Section 5 builds on the bi-directed case to describe a Markov chain Monte Carlo approach for the DMG case. Section 6 describes an alternative variational approach. Experiments are in Section 7.

## 2 REVIEW OF GAUSSIAN MIXED GRAPH MODELS

We describe a common parameterization of acyclic DMGs to represent multivariate Gaussian distributions. This parameterization follows the tradition found in structural equation modeling (Bollen, 1989).

For simplicity of presentation, we will assume through this paper that all random variables have zero mean[1]. For each variable $Y_j$ with parents $Y_{j1}, ..., Y_{jk}$, we provide a "structural equation"

$$Y_j = b_{j1}Y_{j1} + b_{j2}Y_{j2} + \cdots + b_{jk}Y_{jk} + \epsilon_j \qquad (1)$$

where $\epsilon_j$ is a Gaussian random variable with zero mean and variance $v_{jj}$.

Unlike in standard regression models, "error term" $\epsilon_j$ is not necessarily constructed to be independent of parent $Y_p \in \{Y_{j1}, \ldots, Y_{jk}\}$. Instead, the independence is asserted by the graphical structure: $\epsilon_j$ and $Y_p$ are structurally dependent (i.e., irrespective of the parameter values) if $Y_j$ and $Y_p$ are connected by a bi-directed edge. This association is represented by the covariance of $\epsilon_p$ and $\epsilon_j$, $v_{pj}$. The same holds for variables $Y_i$ that are not in the structural equation for $Y_j$ (i.e., not a parent).

By this parameterization, each directed edge $Y_j \to Y_i$ in the graph corresponds to a parameter $b_{ij}$. Each bi-directed edge $Y_i \leftrightarrow Y_j$ in the graph corresponds to parameter $v_{ij}$. Each node $Y_j$ in the graph corresponds to parameter $v_{jj}$. Algebraically, let $\mathbf{B}$ be a $q \times q$ lower triangular matrix, $q$ being the number of observed variables, such that $\mathbf{B}_{ij} = b_{ij}$ if $Y_j \to Y_i$ exists in the graph, and 0 otherwise. Let $\mathbf{V}$ be a $q \times q$ matrix, where $\mathbf{V}_{ij} = v_{ij}$ if $i = j$ or if $Y_i \leftrightarrow Y_j$ is in the graph, and 0 otherwise. Let $\mathbf{Y}$ be the column vector of observed variables, and $\epsilon$ be the corresponding vector of error terms. The set of structural equations is then given in matrix form by

$$\begin{aligned}\mathbf{Y} = \mathbf{B}\mathbf{Y} + \epsilon &\Rightarrow \mathbf{Y} = (\mathbf{I} - \mathbf{B})^{-1}\epsilon \\ \Rightarrow \Sigma(\Theta) &= (\mathbf{I} - \mathbf{B})^{-1}\mathbf{V}(\mathbf{I} - \mathbf{B})^{-T}\end{aligned} \qquad (2)$$

where $\mathbf{A}^{-T}$ is the transpose of the inverse of matrix $\mathbf{A}$ and $\Sigma(\Theta)$ is the *implied covariance matrix* of the model, $\Theta = \{\mathbf{B}, \mathbf{V}\}$.

Other considerations on parameterizing Gaussian mixed graphs are discussed by Richardson and Spirtes (2002). In particular, in the class of Gaussian maximal ancestral graph models, this parameterization will not introduce extra constraints in the joint distribution besides the independence constraints derived by the m-separation criterion.

A common pratice in the structural equation modeling literature is to deal with bi-directed edges (e.g., Dunson et al., 2005) using a different representation: each bi-directed edge $Y_1 \leftrightarrow Y_2$ is replaced by an "ancillary" latent variable with two children only, $Y_1 \leftarrow X \to Y_2$. The resulting DAG is parameterized in the usual way. This, however, might further introduce new constraints besides the independence constraints. Figure 2 illustrates a case where this can happen.

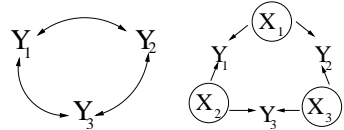

Figure 2: Two representations of hidden common cause associations by using a DMG and a latent variable DAG. Not all covariance matrices can be parameterized with the DAG (Richardson and Spirtes, 2002).

An important advantage of Bayesian inference over maximum likelihood approaches lies on computing posterior distributions for unidentifiable causal effects when informative prior knowledge is available. Such causal effects are functions of parameters that cannot be determined from data alone, i.e., parameters for which there is more than one possible choice of value that generates the same observed covariance matrix.

The simplest case is the "bow-structure" consisting of a DMG of two variables connected by both a directed and a bi-directed edge, e.g., the subgraph of Figure 1 induced by the subset of nodes $\{Y_2, Y_3\}$: there is no unique choice of $v_{23}$ and $b_{32}$ to represent $\sigma_{23}$. There are sophisticated techniques to identify these parameters in several important special cases (e.g., Brito and Pearl, 2002), but in general they might remain unidentifiable. This is not a fundamental problem for Bayesian analysis, provided that an informative prior distribution for parameters is given[2].

## 3 PRIORS

Our prior takes the form $p(\Theta) = p(\mathbf{B})p(\mathbf{V})$. We assign priors for the parameters of directed edges (matrix $\mathbf{B}$) in a standard way: each parameter $b_{ij}$ is given a Gaussian $N(\mu_{ij}^B, \sigma_{ij}^B)$ prior, where all parameters are marginally independent in the prior, i.e., $p(\mathbf{B}) = \prod_{ij} p(b_{ij})$.

To the best of our knowledge, no Bayesian treatment of mixed graphs has been developed before. The difficulty lies on computing Bayesian estimates of models defined on bi-directed graphs (i.e., mixed graphs with bi-directed edges only). To simplify our presentation of priors for $\mathbf{V}$, for the rest of this section we will focus on bi-directed graphs.

---

[1] It is important in Bayesian analysis to represent means explicitly. However, doing that would add much notational complexity in some parts of the paper. The full case is straighforward to derive from the results presented here.

[2] There might be other practical inference problems (particularly with Monte Carlo sampling and large data sets) if, e.g., the posterior surface is flat on large regions.

Gaussian bi-directed graph models are sometimes called *covariance graph models*. Covariance graphs are models of marginal independence: each edge corresponds to a single parameter in the covariance matrix (the corresponding covariance); the absence of an edge $Y_i \leftrightarrow Y_j$ is a statement that $\sigma_{Y_i Y_j} = 0$, $\sigma_{XY}$ being the covariance of random variables $X$ and $Y$. If **W** is a random covariance matrix generated by a covariance model, the distribution of **W** should assign density zero for all matrices with non-zero values on the entries corresponding to non-adjacent nodes.

If $G$ was a fully connected graph, this would reduce to the problem of Bayesian models for unrestricted Gaussians. A common prior for covariance matrices is the inverse Wishart prior $IW(\delta, \mathbf{U})$. In this paper we adopt the following inverse Wishart parameterization:

$$p_{IW}(\Sigma) \propto |\Sigma|^{-(\delta+2q)/2} \exp\left\{-\tfrac{1}{2} tr(\Sigma^{-1}\mathbf{U})\right\}, \quad (3)$$
$$\Sigma \text{ positive definite}$$

with $q$ being the number of variables (nodes) in our model[3].

Following Atay-Kayis and Massam (2005), let $M^+(G)$ be the cone of positive definite matrices such that, for a given bi-directed graph $G$ and $\Sigma \in M^+(G)$, $\Sigma_{ij} = 0$ if nodes $Y_i$ and $Y_j$ are not adjacent in $G$. If one wants a conjugate prior $p(\Sigma)$ for the likelihood function, with mean assumed to be zero and a sufficient statistic matrix $\mathbf{D} = \sum_{i=1}^d (\mathbf{Y}^{(i)})(\mathbf{Y}^{(i)})^T$ calculated from a sample of **Y** of size $d$,

$$p(Data|\Sigma) = (2\pi)^{-dq/2} |\Sigma|^{-d/2} \exp\left\{-\frac{1}{2} tr(\Sigma^{-1}\mathbf{D})\right\} \quad (4)$$

it follows that a possible choice is

$$p(\Sigma) = \frac{1}{I_G(\delta, \mathbf{U})} |\Sigma|^{-(\delta+2q)/2} \exp\left\{-\tfrac{1}{2} tr(\Sigma^{-1}\mathbf{U})\right\},$$
$$\Sigma \in M^+(G) \quad (5)$$

which is basically a re-scaled inverse Wishart prior with a different support and, consequently, different normalizing constant $I_G(\delta, \mathbf{U})$. An analogous concept exists for undirected graphs, where $\Sigma^{-1} \in M^+(G)$ gives a Wishart-like prior: the "$G$-Wishart" distribution (Atay-Kayis and Massam, 2005). We call Equation 5 the *G-Inverse Wishart* prior ($G$-$IW$), which will be the basis of our framework. There are no analytical formulas for the normalizing constant.

One might argue that for computational purposes it is more practical to adopt the "ancillary" latent representation discussed in the previous section: since there is no closed formula posterior for the bi-directed model, one will need some type of sampling procedure or approximation either way. Each covariance term in the ancillary representation can be easily sampled within a Gibbs sampling procedure, a popular practice in Bayesian analysis of causal models with latent variables (Scheines et al., 1999; Dunson et al., 2005), while it is not clear the same can be done with bi-directed parameterizations.

However, we will show in the next section that is possible to do direct Monte Carlo sampling of covariance models defined by bi-directed graphs without resorting to tricks such as importance sampling. In Section 5, we combine it with Gibbs sampling to get a Markov chain Monte Carlo procedure for ADMGs. In Section 7 we show that this procedure can sometimes be much faster than the regular "ancillary" representation while retaining the niceties of the ADMG parameterization.

## 4 A MONTE CARLO ALGORITHM FOR BI-DIRECTED MODELS

The space $M^+(G)$ can be described as the space of positive definite matrices conditioned on the event that each matrix has zero entries corresponding to non-adjacent nodes in graph $G$. In order to do Bayesian inference with Gaussian models on bi-directed graphs, in this section we describe a Monte Carlo procedure to sample from the posterior distribution of covariance matrices with a $G$-$IW$ prior. We follow the framework of Atay-Kayis and Massam (2005) using the techniques of Drton and Richardson (2004).

Atay-Kayis and Massam (2005) show how to sample from a non-decomposable undirected model by reparameterizing the precision matrix throught the Cholesky decomposition. The zero entries in the inverse covariance matrix of this model correspond to constraints in this parameterization, where part of the parameters can be sampled independently and the remaining parameters calculated from the independent ones.

We will follow a similar framework but with a different decomposition. This provides an easy way to sample directly from the $G$-$IW$ distribution.

### 4.1 Bartlett's decomposition

Brown et al. (1993) attribute the following result to Bartlett: a positive definite matrix $\Sigma$, written as the partitioned matrix

$$\Sigma = \begin{pmatrix} \Sigma_{11} & \Sigma_{12} \\ \Sigma_{21} & \Sigma_{22} \end{pmatrix} \quad (6)$$

can be decomposed as $\Sigma = \mathbf{T}\Delta\mathbf{T}^T$ where

---
[3]Although not crucial to our results, we adopt this non-standard parameterization to match the one used by (Brown et al., 1993; Atay-Kayis and Massam, 2005).

$$\Delta = \begin{pmatrix} \Sigma_{11} & 0 \\ 0 & \Gamma \end{pmatrix} \quad \text{and} \quad \mathbf{T} = \begin{pmatrix} \mathbf{I} & 0 \\ \mathcal{B} & \mathbf{I} \end{pmatrix} \quad (7)$$

such that

$$\mathcal{B} = \Sigma_{21}\Sigma_{11}^{-1} \quad \text{and} \quad \Gamma = \Sigma_{22.1} = \Sigma_{22} - \Sigma_{21}\Sigma_{11}^{-1}\Sigma_{12}$$

That is, $\Sigma$ can be parameterized by $(\Sigma_{11}, \mathcal{B}, \Gamma)$ provided a partition $\{\mathbf{Y_1}, \mathbf{Y_2}\}$ of its random variables and the mapping $\Sigma \to \{\Sigma_{11}, \mathcal{B}, \Gamma\}$ is bijective. In this case, $\Sigma_{11}$ is the covariance matrix of $\mathbf{Y_1}$, $\mathcal{B}$ is equivalent to the coefficients obtained by least-squares regression of $\mathbf{Y_2}$ on $\mathbf{Y_1}$, and $\Gamma$ is the covariance matrix of the residuals of this regression. Expressing $\Sigma$ as a function of $\{\Sigma_{11}, \mathcal{B}, \Gamma\}$ gives

$$\Sigma = \begin{pmatrix} \Sigma_{11} & \Sigma_{11}\mathcal{B}^{\mathbf{T}} \\ \mathcal{B}\Sigma_{11} & \Gamma + \mathcal{B}\Sigma_{11}\mathcal{B}^{\mathbf{T}} \end{pmatrix} \quad (8)$$

This decomposition can be applied recursively. Let $\{n\}$ represent the set of indices $\{1, 2, \ldots, n\}$. Let $\Sigma_{i,\{i-1\}}$ be the vector containing the covariance between $Y_i$ and all elements of $\{Y_1, Y_2, \ldots, Y_{i-1}\}$. Let $\Sigma_{\{i-1\},\{i-1\}}$ be the marginal covariance matrix of $\{Y_1, Y_2, \ldots, Y_{i-1}\}$. Let $\sigma_{ii}$ be the variance of $Y_i$. Define the mapping $\Sigma \to \{\gamma_1, \mathcal{B}_2, \gamma_2, \mathcal{B}_3, \gamma_3, \ldots, \mathcal{B}_q, \gamma_q\}$, such that $\mathcal{B}_i$ is a vector with $i-1$ entries, $\gamma_i$ is a scalar, $\sigma_{11} = \gamma_1$, and

$$\begin{aligned} \Sigma_{i,\{i-1\}} &= \mathcal{B}_i \Sigma_{\{i-1\},\{i-1\}}, \ i > 1 \\ \sigma_{ii} &= \gamma_i + \mathcal{B}_i \Sigma_{\{i-1\},i}, \ i > 1 \end{aligned} \quad (9)$$

For a random inverse Wishart matrix, Bartlett's decomposition allows the definition of its density function by the joint density of $\{\gamma_1, \mathcal{B}_2, \gamma_2, \mathcal{B}_3, \gamma_3, \ldots, \mathcal{B}_q, \gamma_q\}$. Define $\mathbf{U}_{\{i-1\},\{i-1\}}$, $\mathbf{U}_{\{i-1\},i}$ and $u_{ii.\{i-1\},\{i-1\}}$ in a way analogous to the $\Sigma$ definitions. The next lemma follows directly from Lemma 1 of Brown et al. (1993):

**Lemma 1** *Suppose $\Sigma$ is distributed as $IW(\delta, \mathbf{U})$. Then, after the transformation $\Sigma \to \Phi = \{\gamma_1, \mathcal{B}_2, \gamma_2, \mathcal{B}_3, \gamma_3, \ldots, \mathcal{B}_q, \gamma_q\}$:*

1. *$\gamma_i$ is independent of $\Phi \setminus \{\gamma_i, \mathcal{B}_i\}$*

2. *$\gamma_i \sim IG((\delta + i - 1)/2, u_{ii.\{i-1,i-1\}}/2)$, where $IG(\alpha, \beta)$ is an inverse gamma distribution*

3. *$\mathcal{B}_i \mid \gamma_i \sim N(\mathbf{U}_{\{i-1\},\{i-1\}}^{-1}\mathbf{U}_{\{i-1\},i}, \gamma_i \mathbf{U}_{\{i-1\},\{i-1\}}^{-1})$, where $N(\mathbf{M}, \mathbf{V})$ is a multivariate Gaussian distribution and $\mathbf{U}_{\{i-1\},\{i-1\}}^{-1} = (\mathbf{U}_{\{i-1\},\{i-1\}})^{-1}$*

## 4.2 Bartlett's decomposition of marginal independence models

What is interesting about Bartlett's decomposition is that it provides a simple parameterization of the inverse Wishart distribution that allows the derivation of new distributions. For instance, Brown et al. (1993) derive a "Generalized Inverted Wishart" distribution that allows one to define different degrees of freedom for different submatrices of an inverse Wishart random matrix. For our purposes, Bartlett's decomposition can be used to reparameterize the $G$-$IW$ distribution.

As pointed out before, $G$-$IW$ is just an inverse Wishart distribution conditioned on the fact that some entries are identically zero. To impose the constraint that $Y_i$ is uncorrelated with $Y_j$, $i > j$, is to set $(\mathcal{B}_i \Sigma_{\{i-1\},\{i-1\}})_j = \sigma_{Y_i Y_j}(\Phi) = 0$. For a fixed $\Sigma_{\{i-1\},\{i-1\}}$, this implies a constraint on $(\mathcal{B}_i)_j \equiv \beta_{ij}$.

This provides a way of sampling covariance matrices in the $G$-$IW$ distribution. Using the parameterization $\Sigma \to (\gamma_1, \mathcal{B}_2, \gamma_2, \mathcal{B}_3, \gamma_3, \ldots, \mathcal{B}_q, \gamma_q)$, $q$ being the number of variables, one samples $\gamma_1$ and then the $(\mathcal{B}_i, \gamma_i)$ parameters following the order $i = 2, 3, \ldots, q$. The constraint is, when sampling the vector $\mathcal{B}_i$, some of its elements will be functions of other elements and the sampled matrix $\Sigma_{\{i-1\},\{i-1\}}$. This is done as follows.

Following the terminology used by Richardson and Spirtes (2002), let a *spouse* of node $Y$ in a mixed graph be any node adjacent to $Y$ by a bi-directed edge. The set of spouses of $Y_i$ is denoted by $sp(i)$. The set of spouses of $Y_i$ *according to order* $Y_1, Y_2, \ldots, Y_q$ is defined by $sp_{\prec}(i) = sp(i) \cap \{Y_1, \ldots, Y_{i-1}\}$. The set of non-spouses of $Y_i$ is denoted by $nsp(i)$. Analogously, $nsp_{\prec}(i) = \{Y_1, \ldots, Y_{i-1}\} \setminus sp_{\prec}(i)$. Let $\mathcal{B}_{i,sp_{\prec}(i)}$ be the subvector of $\mathcal{B}_i$ corresponding to the "regression" coefficients of $Y_i$ on its ordered spouses. Let $\mathcal{B}_{i,nsp_{\prec}(i)}$ be the complementary vector.

To sample $\mathcal{B}_i$ from a $G$-$IW$, one just have to condition on the event $(\mathcal{B}_i \Sigma_{\{i-1\},\{i-1\}})_j = \sigma_{Y_i Y_j} = 0$ for all nodes $Y_j$ such that $Y_j \in nsp_{\prec}(i)$. This can be achieved by first sampling $\mathcal{B}_{i,sp_{\prec}(i)}$ from the respective marginal of the multivariate Gaussian given by Lemma 1. From the identity $\mathcal{B}_i \Sigma_{\{i-1\},nsp_{\prec}(i)} = 0$, it follows

$$\mathcal{B}_{i,sp_{\prec}(i)} \Sigma_{sp_{\prec}(i),nsp_{\prec}(i)} + \mathcal{B}_{i,nsp_{\prec}(i)} \Sigma_{nsp_{\prec}(i),nsp_{\prec}(i)} = 0 \quad (10)$$

$$\mathcal{B}_{i,nsp_{\prec}(i)} = -\mathcal{B}_{i,sp_{\prec}(i)} \Sigma_{sp_{\prec}(i),nsp_{\prec}(i)} \Sigma_{nsp_{\prec}(i),nsp_{\prec}(i)}^{-1} \quad (11)$$

Those identities are also derived by Drton and Richardson (2004) under the context of maximum likelihood estimation. In that case, the estimates for the elements of the covariance matrix are in general coupled. Therefore, their procedure is an iterative algo-

**Algorithm** SAMPLEGINVERSEWISHART
Input: matrix $\mathbf{K}$, scalar $n$, bi-directed graph $G$

1. Let $\Sigma$ be a $q \times q$ matrix, with $q$ being the number of rows in $\mathbf{K}$
2. Define functions $sp_\prec(i)$, $nsp_\prec(i)$ according to $G$ and ordering $Y_1, \ldots, Y_q$
3. Sample $\sigma_{11}$ from $IG(n/2, k_{11}/2)$
4. For $i = 2, 3, \ldots, q$
5.     Sample $\gamma_i \sim IG((n+i-1)/2, \mathbf{K}_{ii.\{i-1\},\{i-1\}})$
6.     Let $\mathbf{M}_i = \mathbf{K}_{\{i-1\},\{i-1\}}^{-1} \mathbf{K}_{\{i-1\},i}$
7.     Sample $\mathcal{B}_{i,sp_\prec(i)}$ from the corresponding marginal of $N(\mathbf{M}_i, \gamma_i \mathbf{K}_{\{i-1\},\{i-1\}}^{-1})$
8.     Set $\mathcal{B}_{i,nsp_\prec(i)} = -\mathcal{B}_{i,sp_\prec(i)} \Sigma_{sp_\prec(i),nsp_\prec(i)} \Sigma_{nsp_\prec(i),nsp_\prec(i)}^{-1}$
9.     Set $\Sigma_{i,\{i-1\}}^T = \Sigma_{\{i-1\},i} = \Sigma_{\{i-1\},\{i-1\}} \mathcal{B}_i$
10.    Set $\sigma_{ii} = \gamma_i + \Sigma_{i,\{i-1\}} \mathcal{B}_i$
11. Return $\Sigma$.

Figure 3: A procedure for sampling from a $G$-Inverse Wishart distribution.

rithm where, for each variable $Y_i$, a constrained regression is performed by conditioning on all other variables and the current estimate of the covariance matrix. Notice that our Monte Carlo method uses an order for conditioning such that $Y_i$ is regressed only on $Y_1, \ldots, Y_{i-1}$.

Every time we sample a pair of parameters $\{\gamma_i, \mathcal{B}_i\}$, we construct the corresponding marginal covariance matrix of $\{Y_1, \ldots, Y_i\}$, which will be used to constrain the next entries of the covariance matrix. The full algorithm is given in Figure 3. Notice that if one wants to sample from the posterior given by prior parameters $(\delta, \mathbf{U})$ and sample sufficient statistics $\mathbf{D} = \sum_{i=1}^{d} (\mathbf{Y}^{(i)})(\mathbf{Y}^{(i)})^T$, the conjugacy property of $G$-IW implies that we just need to sample from a $G$-$IW(\delta + d, \mathbf{U} + \mathbf{D})$ distribution.

### 4.3 The normalizing constant

Bayesian modeling of covariance graphs is also useful for model selection. By combining the likelihood equation (4) with the prior (5), we obtain the joint

$$p(\mathbf{D}, \Sigma | G) = (2\pi)^{-\frac{dq}{2}} I_G(\delta, \mathbf{U})^{-1}$$
$$\times |\Sigma|^{-\frac{\delta + 2q + d}{2}} \exp\left\{-\frac{1}{2} tr[\Sigma^{-1}(\mathbf{D} + \mathbf{U})]\right\} \quad (12)$$

where we make the dependency on the graphical structure $G$ now explicit. By the definition of $I_G$, integrating $\Sigma$ out of the above equation implies the following marginal likelihood:

$$p(\mathbf{D}|G) = \frac{1}{(2\pi)^{\frac{dq}{2}}} \frac{I_G(\delta + d, \mathbf{D} + \mathbf{U})}{I_G(\delta, \mathbf{U})} \quad (13)$$

Therefore, using the marginal likelihood for model selection (as in, e.g., given uniform priors for $G$) requires the computation of integrals of the type

$$I_G(n, \mathbf{K}) = \int_{M^+(G)} |\Sigma|^{-\frac{n+2q}{2}} \exp\left\{-\frac{1}{2} tr(\Sigma^{-1} \mathbf{K})\right\} d\Sigma \quad (14)$$

We will use Bartlett's decomposition again for the case where $G$ is not a complete graph. Notice from Section 4.2 that there is a one-to-one correspondence between each edge $Y_i \leftrightarrow Y_j$, $i < j$ and free parameter $\beta_{ji} \in \mathcal{B}_j$. There is also a one-to-one correspondence between each pair of non-adjacent nodes $\{Y_k, Y_j\}$, $k < j$, and the constrained parameter $\beta_{jk} \in \mathcal{B}_j$ that is a function of the free parameters.

Let $\Phi^E$ be the set of free parameters, corresponding to $\{\gamma_1, \ldots, \gamma_q\}$ and the elements of $\mathcal{B}$ associated with the edges in the graph. The mapping between entries in $\Sigma$ and $\Phi^E$ is bijective and differentiable. The determinant of the Jacobian of this transformation is given by the following lemma:

**Lemma 2** *The determinant of the Jacobian for the change of variable* $\Sigma \rightarrow \Phi^E$ *is*

$$|J(\Phi^E)| = \prod_{i=1}^{q-1} \gamma_i^{\#sp_\succ(i)} \quad (15)$$

*where* $\#sp_\succ(i)$ *is the number of elements in* $sp(i) \cap \{Y_{i+1}, Y_{i+2}, \ldots, Y_q\}$.

**Proof sketch:** Represent $\Sigma$ as the column vector $\Sigma = [\sigma_{11}, \sigma_{21}, \sigma_{22}, \sigma_{31}, \sigma_{32}, \sigma_{33}, \ldots, \sigma_{qq}]^T$, but excluding those $\sigma_{ji}$ entries which are identically zero by construction. Represent $\Phi$ as the row vector $\Phi = [\gamma_1, \beta_{21}, \gamma_2, \beta_{31}, \beta_{32}, \ldots, \gamma_q]$. Vector $\Phi^E$ is given by the row vector $\Phi$, but excluding those $\beta_{ji}$ that are not free (i.e., that do not correspond to any edge $Y_i \leftrightarrow Y_j$). Therefore, the $i$th row of matrix $\partial(\Sigma)/\partial(\Phi^E)$ is the gradient of the $i$th element of $\Sigma$ with respect to $\Phi^E$, where only non-zero, non-repeated elements of $\Sigma$ are considered, following the specified order.

Notice that $\partial \sigma_{ji}/\partial \beta_{st} = 0$ and $\partial \sigma_{ji}/\partial \gamma_s = 0$ for $s > j$ (by construction, $j \geq i$ and $s \geq t$). This implies that $J(\Phi^E)$ is a block (lower) triangular matrix of $q$ blocks. A few lines of algebra show that desired determinant is given by the determinant of the blocks

$$|J(\Phi^E)| = \prod_{i=1}^{q} |\Sigma_{sp_\prec(i),sp_\prec(i)}| \quad (16)$$

where $\Sigma_{sp_\prec(i),sp_\prec(i)}$ is the marginal covariance matrix of the preceding spouses of $Y_i$ according to $\prec$. One can show that $|\Sigma_{sp_\prec(i),sp_\prec(i)}| = |\Gamma_{sp_\prec(i)}| = \prod_{v \in sp_\prec(i)} \gamma_v$, where $\Gamma_{sp_\prec(i)}$ is a diagonal matrix with diagonal given by the $\gamma$ parameters associated with $sp_\prec(i)$. By combining this with (16), the result follows. □

The expression for the normalizing constant $\delta(n, \mathbf{K})$ is then given by

$$\begin{aligned} I_G(n, \mathbf{K}) &= \int |J(\Phi^E)||\Sigma(\Phi^E)|^{-\frac{n+2q}{2}} \\ &\times \exp\left\{-\frac{1}{2}tr(\Sigma(\Phi^E)^{-1}\mathbf{K})\right\} d\Phi^E \end{aligned} \quad (17)$$

Let $\langle f(X) \rangle_{p(X)}$ be the expected value of $f(X)$ with respect to distribution $p(X)$. Let $p_{IG}(\gamma|\alpha, \beta)$ be the inverse gamma density function with parameters $\{\alpha, \beta\}$. Let $p_N(\mathcal{B}|\mathbf{M}, \mathbf{C})$ be a multivariate Gaussian density function with mean $\mathbf{M}$ and covariance matrix $\mathbf{C}$. Let $p(\Phi^E)$ be the joint distribution for the free parameters in our model defined by Bartlett's decomposition and equal to the respective product of inverse gamma and multivariate normal distributions defined by it. The following theorem restates integral (17):

**Theorem 1** *The normalizing constant of a G-Inverse Wishart for matrix $\Sigma$ with parameters $(n, \mathbf{K})$ is given by*

$$\begin{aligned} I_G(n, \mathbf{K}) &= I_{IW}(n, \mathbf{K}) \\ &\times \left\langle \prod_{i=1}^q \gamma_i^{\#sp_\succ(i)-(q-i)} \prod_{i=2}^q f_i(\Phi^E) \right\rangle_{p(\Phi^E)} \end{aligned} \quad (18)$$

*where $I_{IW}(n, \mathbf{K})$ is the normalizing constant of an inverse Wishart with $n + q - 1$ degrees of freedom and matrix parameter $\mathbf{K}$.*

*Moreover, $f_i(\Phi^E)$ corresponds to a Gaussian density function $p_N(\mathcal{B}_{i,sp_\prec(i)}|\mathbf{M}, \mathbf{C})$ where: $\mathcal{B}_{i,sp_\prec(i)}$ equals the expression in Equation (11), which is a function of $\Phi^E$; $\mathbf{M}$ and $\mathbf{C}$ correspond, respectively, to the mean and covariance of the distribution of $\mathcal{B}_{i,nsp_\prec(i)}|\mathcal{B}_{i,sp_\prec(i)}$. If $nsp_\prec(i) = \emptyset$, $f_i(\Phi^E)$ is defined to be 1.*

**Proof sketch:** In an inverse Wishart distribution, the equivalent to integral (17) is the integral over $I_{IW}(n, \mathbf{K}) \times p_{IW}(n, \mathbf{K})$, the last factor being the density function for an inverse Wishart. By Lemma 1, the density function $p_{IW}(n, \mathbf{K})$ of a graphical inverse Wishart model with all edges, parameterized by $\Phi$, can be represented by the product on $q$ inverse gamma density functions and $q - 1$ multivariate normal functions. To obtain the equivalent expression with respect to $\Phi^E$, $E$ being the set of edges, we just need to divide the exactly same product by $|J(\Phi)|$ and multiply it by $|J(\Phi^E)|$. By Lemma 2, $|J(\Phi^E)|/|J(\Phi)| = \prod_{i=1}^q \gamma_i^{\#sp_\succ(i)-(q-i)}$. The rest follows by splitting the density function of $p_N(\mathcal{B}_i)$ into the marginal for $\mathcal{B}_{i,sp_\prec(i)}$ (incorporated into $p(\Phi^E)$), and conditional $p_N(\mathcal{B}_{i,nsp_\prec(i)}|\mathcal{B}_{i,sp_\prec(i)})$ evaluated at

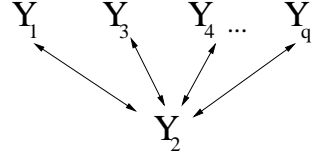

Figure 4: The computational cost of sampling from the model represented above will vary depending on the position of $Y_2$ in the chosen order.

the values given by Equation (11) (i.e., $f_i(\Phi^E)$). □

Samples $\{\Phi^{E(1)}, \Phi^{E(2)}, \ldots, \Phi^{E(m)}\}$ can be generated according to the algorithm given in Figure 3. A Monte Carlo estimate of $I_G(n, \mathbf{K})$ is given by:

$$I_G(n, \mathbf{K}) \approx \frac{I_{IW}(n, \mathbf{K})}{m} \sum_{s=1}^m g(\Phi^{E(s)}) \quad (19)$$

where $g(\Phi^E) = \prod_{i=1}^q \gamma_i^{\#sp_\succ(i)-(q-1)} \prod_{i=2}^q f_i(\Phi^E)$.

## 5 GIBBS SAMPLING FOR ADMGs

To extend our method to directed mixed graphs with parameters $\Theta = \{\mathbf{V}, \mathbf{B}\}$, we apply a Gibbs sampling procedure that alternates sampling $\mathbf{V}$ given $\mathbf{B}$ and vice-versa.

The conditional distribution for $\mathbf{V}$, the covariance matrix for error terms, is given by

$$p(\mathbf{V}|\mathbf{D}, \mathbf{B}) = G\text{-}IW(\delta + d, \mathbf{U} + (\mathbf{I} - \mathbf{B})\mathbf{D}(\mathbf{I} - \mathbf{B})^T) \quad (20)$$

which can be sampled by using the algorithm described in the previous section. The sufficient statistic matrix $\mathbf{D}$ can trivially include sampled latent variables.

Conditional distributions for $\mathbf{B}$ given $\mathbf{V}$ can be derived as in previous approaches for Gaussian DAGs (Dunson et al., 2005; Scheines et al., 1999).

Factoring the conditional distributions can speed up the algorithm. While sampling covariance matrices $\mathbf{V}$, it is possible to sample some entries independently. To achieve a reasonable factorization, it is of interest to choose an order that requires fewer and smaller matrix inversions. Matrix inversions are necessary when regressing $Y_i$ on $\{Y_1, \ldots, Y_{i-1}\}$.

For instance, consider the graph shown in Figure 4. Order $\{Y_1, Y_2, Y_3, \ldots, Y_q\}$ will require $q - 1$ inversions of matrices of size $\{1, 2, \ldots, q - 1\}$, respectively, since all nodes $Y_i$, $i > 3$ will be dependent on all previous nodes given $Y_2$. In constrast, order

$\{Y_1, Y_3, Y_4, \ldots, Y_q, Y_2\}$ will require only one matrix inversion ($Y_2$ regressed on all other nodes).

It is also possible to decompose the conditional distribution for the parameters of directed edges up to some extent. In the extreme, if the graph is a DAG (i.e., no bi-directed edges), then distribution $p(\mathbf{B}|\mathbf{V}, \mathbf{D})$ factorizes into $q$ factors, where each factor is composed only of parameters associated with edges into a single node. This may lead to much smaller matrices that have to be inverted in the derivation of this posterior. The factorization will depend on the bi-directed edges. We omit the proof due to space constraints, but one can show that the conditional distribution for $\mathbf{B}$ cannot in general be factorized beyond the set of *districts* of the ADMG, as defined by Richardson (2003).

# 6 A VARIATIONAL MONTE CARLO APPROXIMATION

The Gibbs sampling procedure can be computationally expensive, especially if the dimensionality of the problem is high or if the goal is automated model selection. In this section, we describe a variational approximation for computing the marginal likelihood of a mixed graph representation. This approximation still makes use of the Monte Carlo sampler for covariance graph models. We adopt the following approximation in our variational approach, allowing also for latents $\mathbf{X}$:

$$p(\mathbf{V}, \mathbf{B}, \mathbf{X}|\mathbf{Y}) \approx q(\mathbf{V})q(\mathbf{B}) \prod_{i=1}^{d} q(\mathbf{X}^{(i)}) = q(\mathbf{V})q(\mathbf{B})q(\mathbf{X})$$

where $q(\mathbf{B})$ and $q(\mathbf{X}^{(i)})$ are multivariate Gaussian and $q(\mathbf{V})$ is a $G$-Inverse Wishart.

From Jensen's inequality, we obtain the following lower-bound (Beal, 2003, p. 47):

$$\begin{aligned}
\ln p(\mathbf{Y}) &= \ln \int p(\mathbf{Y}, \mathbf{X}|\mathbf{V}, \mathbf{B}) p(\mathbf{V}, \mathbf{B}) \, d\mathbf{X} d\mathbf{B} d\mathbf{V} \\
&\geq \langle \ln p(\mathbf{Y}, \mathbf{X}|\mathbf{V}, \mathbf{B}) \rangle_{q(\mathbf{V})q(\mathbf{B})q(\mathbf{X})} \\
&\quad + \langle \ln p(\mathbf{V})/q(\mathbf{V}) \rangle_{q(\mathbf{V})} \\
&\quad + \langle \ln p(\mathbf{B})/q(\mathbf{B}) \rangle_{q(\mathbf{B})} - \langle \ln q(\mathbf{X}) \rangle_{q(\mathbf{X})}
\end{aligned}$$

where this lower bound can be optimized with respect to functions $q(\mathbf{V})$, $q(\mathbf{B})$, $q(\mathbf{X})$. This can be done by iterative coordinate ascent, maximizing the bound with respect to a single $q(\cdot)$ function at a time.

The update of $q(\mathbf{V})$ is given by

$$q^{new}(\mathbf{V}) = p_{G\text{-}IW}(\delta+d, \mathbf{U} + \langle (\mathbf{I}-\mathbf{B})\mathbf{D}(\mathbf{I}-\mathbf{B})^T \rangle_{q(\mathbf{X})q(\mathbf{B})})$$

where $\mathbf{D}$ is the empirical second moment matrix summed over the completed dataset $(\mathbf{X}, \mathbf{Y})$. The updates for $q(\mathbf{B})$ and $q(\mathbf{X})$ are tedious but straighforward derivations, omitted for space purposes. The relevant fact about these updates is that they are functions of $\langle \mathbf{V}^{-1} \rangle_{q(\mathbf{V})}$. Fortunately, one pays a small cost to obtain these inverses from the Monte Carlo sampler of Figure 3: from Bartlett's decomposition, one can create a lower triangular matrix $\mathcal{B}$ (by placing on the $i$th line the row vector $\mathcal{B}_i^T$, followed by zeroes) and a diagonal matrix $\Gamma$ from the respective vector of $\gamma_i$'s. A sample of $\mathbf{V}^{-1}$ can be obtained from $(\mathbf{I}-\mathcal{B})^T \Gamma^{-1} (\mathbf{I}-\mathcal{B})$.

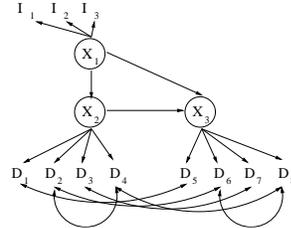

Figure 5: A mixed graph for the industrialization and democratization domain.

# 7 EXPERIMENT

We now evaluate how Markov Chain Monte Carlo computation on DAGs, where bi-directed edges are substituted by latent variables (DAG-MCMC), compares to our directed mixed graph MCMC (ADMG-MCMC) in computational time. We also evaluate the quality of our variational Monte Carlo bound by comparing its predictive log-likelihood against Gibbs sampling[4].

The dataset we will use is a study of the relationship between democratization and industrialization in developing countries. This is a longitudinal study containing indicators of 1960 and 1965. The data contains 75 samples and 11 variables. It contains three indicators of latent variable "industrialization" ($I_1, I_2, I_3$), four indicators of latent variable "democracy" in 1960 ($D_1, \ldots, D_4$) and in 1965 ($D_5, \ldots, D_8$). A mixed graph for this domain (Figure 5), priors and other details are given by Bollen (1989) and Dunson et al. (2005).

Figure 6 illustrates some of the properties of our sampler concerning different aspects, such as convergence of sampling and of the posterior mean estimator. The last graph in Figure 6 shows a comparison of the our MCMC sampler and the variational MC approach regarding predictive log-likelihood averaged over test points. The figure shows the outcomes out of a 10-fold cross-validation study[5]. The result shown in Figure 6

---
[4]DAG-MCMC and ADMG-MCMC give approximately the same predictive log-likelihood in this study. This is expected when the graph does not contain large chains or cycles of bi-directed edges, since the family of models encoded by the DAG is approximately the same.

[5]We use the method described by Beal (2003), §4.4, to

shows that in this dataset there is no statistically significant difference (t-test, p-value = 0.05) between the variational MC and the MCMC regarding predictions.

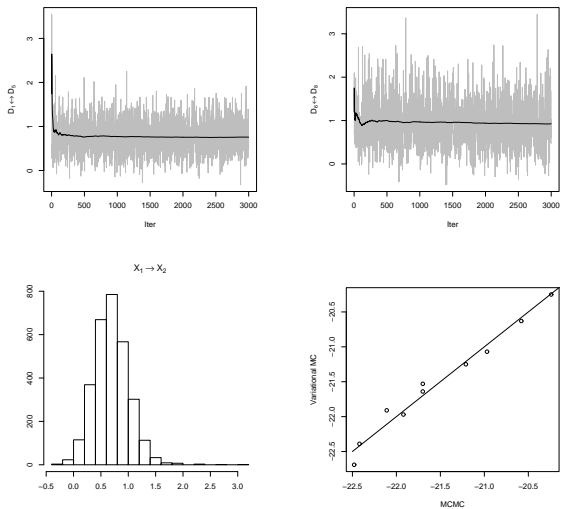

Figure 6: An illustration of the behaviour of the MCMC algorithm. Top row: convergence of the sampler for the parameters associated with $D_1 \leftrightarrow D_5$ and $D_6 \leftrightarrow D_8$. The solid line represents the posterior mean estimator across iterations. Botton left: histogram of the posterior for the parameter associated with $X_1 \rightarrow X_2$, comparable to the results given by Dunson et al. (2005). Bottom right: the average predictive log-likelihood of the MCMC against the variational procedure in a 10-fold cross-validation.

We also ran our ADMG sampler and the DAG-MCMC sampler for 5000 iterations[6] and compared the computational cost. To be fair to DAG-MCMC, we fix the value of the parameter associated with one of the edges out of each ancillary latent. We do not further fix the second edge nor the value of the variance of the ancillary latent, since this might add too many constraints to the implied distribution[7]. All latents are sampled conditioned only on their respective Markov blankets. By averaging over 10 trials in a Pentium IV 1.8Ghz running on Java, the ADMG-MCMC sampler took 55.7 seconds (std. dev. of 1.8 secs) and the DAG-

---

calculate the variational predictive log-likelihood.

[6] By visual inspection, DAG-MCMC seems to require at least as many samples to converge as ADMG-MCMC in this study, unsurprising since DAG-MCMC has to deal with many more latents. Hence, our comparison here is conservative. The variational approximation is not computationally advantageous for this small dataset, but we expect it to be reasonably faster for larger domains.

[7] Dunson et al. (2005) fix the parameters of both edges, which will fix the sign of the correlation due to bi-directed edges. This is acceptable in this domain, but not in general.

MCMC sampler took 110.9 seconds (dev. 2.1 secs).

To summarize, as compared to DAGs, the ADMG approach provides not only a more direct representation of conditional independencies, but we found that ADMG sampling was more efficient in our experiment.

## 8 CONCLUSIONS

We presented a solution to the problem of Bayesian modeling of Gaussian mixed graph models. Notice that it could be combined with the work on undirected models of Atay-Kayis and Massam (2005) for graphs that include undirected edges. Future possibilities include the design of more efficient algorithms for special types of mixed graphs such as ancestral graphs (Richardson and Spirtes, 2002) and for other families of probabilistic models that use Gaussian models as a component, e.g. latent trait models for ordinal data.

### Acknowledgements

We thank Kenneth Bollen for providing us with the democratization and industrialization data. This work was funded by the Gatsby Charitable Foundation.